\documentclass[preprint,showpacs,showkeys,preprintnumbers,amsmath,amssymb]{revtex4}
 \usepackage{dcolumn}
 \usepackage{bm}
\usepackage{graphicx}
 \begin{document}

 \title{ Stability of charged black holes in string theory under
 charged massive scalar perturbations }

 \author{Ran Li}

 \thanks{Electronic mail: liran.gm.1983@gmail.com}

 \affiliation{Department of Physics,
 Henan Normal University, Xinxiang 453007, China}

 \begin{abstract}

 Similar to the superradiant effect in Reissner-Nordstr\"{o}m black hole,
 a charged scalar field can be amplified when impinging on the charged
 black hole in string theory. According to the black-hole bomb mechanism,
 the mass term of the incident field can works as the
 reflecting mirror, which may trigger the instability of black hole.
 We study the possible instability triggered
 by superradiant effect and demonstrate that
 the charged black hole in string theory is
 stable against the massive charged scalar perturbation. The reason is
 that there is no trapping potential well in the black hole exterior
 which is separated from the horizon by a potential barrier
 and there is no bound states in the superradiant regime.

 \end{abstract}



 \maketitle

 The stability tests of black holes become an important topics
 since the initial work of Regge and Wheeler \cite{reggewheeler}.
 The Schwarzschild black hole is well known to be stable
 \cite{reggewheeler,schsta}. The studying of stability for
 the charged or the rotating black holes becomes involved
 because of the significant effect of classical black holes,
 i.e. superradiance. If an incident bosonic wave whose frequency
 lies in the superradiant regime, the wave scattered by the event
 horizon get amplified \cite{zeldovich,bardeen,misner,starobinsky}.
 This classical effect allows extracting rotational
 energy efficiently from the rotating black holes or extracting
 the Coulomb energy from charged black holes \cite{bekenstein}. The stability analysis
 of these types of black holes should be adequately addressed.

 Press and Teukolsky \cite{press} proposed to use the superradiance to built
 the black-hole bomb by adding an reflecting mirror outside the black hole
 (see also \cite{cardoso2004bomb,Hod,Rosa,Lee,PCGBI,herdeiro,hod2013prd}
 for a recent studies on this topic).
 In this mechanism, the amplitudes of superradiant modes trapped in the potential well
 between the mirror and the event horizon will grow exponential.
 Later, it is found that the mass term of the scalar field or the boundary of
 anti de-sitter (AdS) spacetime can play the roll of the reflecting mirror
 \cite{kerrunstable,detweiler,dolan,hodPLB2012,konoplyaPLB,DiasPRD2006,
 WCIS,dolanprd2013,BCP,jgrosa,cardoso2004ads,cardoso2006prd, murata,KKZ,aliev,uchikata,clement,knopolya,rlplb,rlepjc,zhang,zhangw}.
 In these cases, the superradiance of the scalar field perturbation will
 lead to the instability of the black hole.

 Recently, it is proved by Hod in \cite{hodrnplb2012,hodrnplb2013} that
 the existence of a trapping potential well outside the black hole
 and superradiant amplification of the trapped modes cannot be satisfied simultaneously.
 This means that the Reissner-Nordstr\"{o}m (RN) black holes
 are stable under the perturbations of massive charged scalar fields.
 However, whether all of the charged black holes are stable is still an open question worth
 studying.

 In this work, we will study the stability
 of the charged black hole in string theory.
 Similar to the superradiant effect in Reissner-Nordstr\"{o}m black hole,
 a charged scalar field can be amplified when impinging on the charged
 black hole in string theory. According to the black-hole bomb mechanism,
 the mass term of the incident field can effectively works as the
 reflecting mirror, which may trigger the instability of black hole.
 However, we will demonstrate that
 the charged black hole in string theory is
 stable against the massive charged scalar perturbation.
 By carefully studying the behavior of the effective potential
 outside the horizon, we will show that there is no trapping potential well
 in the black hole exterior which is separated from the horizon
 by a potential barrier. So there is no bound states
 in the superradiant regime which can leat to the instability
 of charged stringy black hole.

 We consider the static charged black holes in low energy effective theory
 of heterotic string theory in four dimensions. Besides the Einstein-Hilbert
 term, the action also includes the contributions from the Dilaton field and the
 Maxwell's field. The charged black hole in string theory
 which is firstly found by Gibbons and Maeda in \cite{GM} and independently
 found by Garfinkle, Horowitz, and strminger in \cite{GHS} a few years later
 is described by the metric
 \begin{eqnarray}
 ds^2=-\left(1-\frac{2M}{r}\right)dt^2+\left(1-\frac{2M}{r}\right)^{-1}
 dr^2+r\left(r-\frac{Q^2}{M}\right)(d\theta^2+\sin^2\theta d\phi^2)\;,
 \end{eqnarray}
 and the electric field and the dialton field
 \begin{eqnarray}
 A_t=-\frac{Q}{r}\;,\;\;\;e^{2\Phi}=1-\frac{Q^2}{Mr}\;.
 \end{eqnarray}

 The parameters $M$ and $Q$ are the mass and electric charge
 of the black hole respectively.
 The event horizon of black hole is located at $r=2M$.
 The area of the sphere of the charged
 stringy black hole approaches to zero when $r=Q^2/M$. Therefore, the
 sphere surface of the radius $r=Q^2/M$ is singular. When $Q^2\leq 2M^2$,
 this singular surface is surrounded by the event horizon.
 We will consider the black hole with the parameters
 satisfying the condition $Q^2\leq 2M^2$ in this paper.
 When $Q^2=2M^2$, the singular surface coincides with the
 event horizon. This is the case of extremal black hole.

 We start with analysing the scalar field perturbation in the background
 of the charged stringy black hole.
 The dynamics of the charged massive scalar field perturbation is governed
 by the Klein-Gordon equation
 \begin{eqnarray}
 \left[(\nabla_\nu-iqA_\nu)(\nabla^\nu-iqA^\nu)-\mu^2\right]\Psi=0\;,
 \end{eqnarray}
 where $q$ and $\mu$ denote the charge and the mass of the scalar field.
 By taking the ansatz of the scalar field
 $\Psi=e^{-i\omega t}R(r)Y_{lm}(\theta,\phi)$,
 where $\omega$ is the conserved energy of the mode,
 $l$ is the spherical harmonic index, and $m$ is the
 azimuthal harmonic index with $-l\leq k\leq l$,
 one can deduce the radial wave equation in the form of
 \begin{eqnarray}
 \Delta\frac{d}{dr}\left(\Delta \frac{dR}{dr}\right)+UR=0\;,
 \end{eqnarray}
 where we have introduced a new function $\Delta=\left(r-Q^2/M\right)\left(r-2M\right)$
 and the potential function is given by
 \begin{eqnarray}
 U=\left(r-\frac{Q^2}{M}\right)^2(\omega r-qQ)^2-\Delta\left[\mu^2r\left(r-\frac{Q^2}{M}\right)+l(l+1)\right]\;.
 \end{eqnarray}

 Firstly, we want to demonstrate that the classical superradiance
 is present for scalar field perturbation in charged
 stringy black hole. In particular, we shall study the asymptotic
 solutions of the radial wave equation near the horizon and
 at spatial infinity with the appropriate boundary conditions
 and obtain the superradiant condition of the charged scalar field.

 To this end, it is convenient to define the tortoise coordinate
 $r_*$ by the equation $dr_*/dr=r^2/\Delta$,
 and introduce the new radial function as
 $\tilde{R}=rR(r)$.
 The domain of the tortoise coordinate $r_*$ is $(-\infty, +\infty)$
 corresponding to the domain $(2M, +\infty)$ of coordinate $r$.
 In terms of the tortoise coordinate and the new radial function,
 the radial wave equation can be rewritten as
 \begin{eqnarray}
 \frac{d^2\tilde{R}}{dr_*^2}+\tilde{U}\tilde{R}=0\;,
 \end{eqnarray}
 with
 \begin{eqnarray}
 \tilde{U}=\frac{U}{r^4}-\frac{\Delta}{r^3}
 \frac{d}{dr}\left(\frac{\Delta}{r^2}\right)\;.
 \end{eqnarray}
 It can be easily checked that when $Q=0$ the above equation
 reduced to the Regge-Wheeler equation \cite{reggewheeler}
 as expected.

 It is easy to obtain the asymptotic behavior of the potential
 $\tilde{U}$ as
 \begin{eqnarray}
 &&\lim_{r\rightarrow 2M}\tilde{U}=\left(1-\frac{Q^2}{2M^2}\right)^2
 \left(\omega-\frac{qQ}{2M}\right)^2\;,\\
 &&\lim_{r\rightarrow \infty}\tilde{U}=\omega^2-\mu^2\;.
 \end{eqnarray}
 So, the radial wave equation has the following asymptotic behavior
 when $\omega^2>\mu^2$
 \begin{eqnarray}
 \tilde{R}=\left\{
             \begin{array}{ll}
               e^{i\sigma r_*}+Ae^{-i\sigma r_*}\;,\;\; r\rightarrow 2M\;,\\
               Be^{i\sqrt{\omega^2-\mu^2}r_*}\;,\;\;\; r\rightarrow \infty\;,
             \end{array}
           \right.
 \end{eqnarray}
 with the parameter $\sigma=\left(1-\frac{Q^2}{2M^2}\right)
 \left(\omega-\frac{qQ}{2M}\right)$. By calculating the radial flux
 of the scalar field modes respectively, one can see that
 the modes of the form $e^{i\sigma r_*}$ and $e^{-i\sigma r_*}$ are outcoming
 from the event horizon and ingoing to the event horizon respectively,
 while the modes of the form $e^{i\sqrt{\omega^2-\mu^2}r_*}$ correspond to
 the outgoing flux at the spatial infinity.
 This boundary conditions then correspond to the flux coming from the
 event horizon of the black hole which is partially reflected back to the
 black hole and partially sent to the spatial infinity.

 Given any two linearly independent solutions $\varphi_1(r_*)$ and $\varphi_2(r_*)$ of
 the radial equation (10), their Wroskian is constant, i.e.
 \begin{eqnarray}
 W(\varphi_1,\varphi_2)\equiv \varphi_1\frac{d\varphi_2}{dr_*}
 -\varphi_2\frac{d\varphi_1}{dr_*}=\textrm{constant}
 \end{eqnarray}
 Because the potential $\tilde{U}$ is real, the asymptotic solution (14) and its
 complex conjugate are linearly independent to each other. Evaluating the Wronskian
 for the asymptotic solution (14) and its complex conjugate
 at the horizon and spatial infinity respectively, we can get
 \begin{eqnarray}
 \sigma(1-|A|^2)=\sqrt{\omega^2-\mu^2}|B|^2\;.
 \end{eqnarray}
 Now, we can see that if $\sigma<0$, we have $|A|^2>1$.
 This means that an incident wave from the horizon is
 reflected back to the horizon with an increased amplitude.
 This phenomenon is known as superradiance.
 From $\sigma<0$, we can get the condition to occur the superradiance
 \begin{eqnarray}
 \omega<q\Phi_H\;,
 \end{eqnarray}
 with $\Phi_H=\frac{Q}{2M}$ being the electric potential at the horizon.
 In fact, this condition has already been obtained in \cite{dilatonsr},
 where the superradiant effects for the dilaton black holes
 are studied.

 Now we shall analysis whether this superradiant effect
 will lead to the instability of the charged stringy black hole.
 The purpose can be achieved by studying whether there exists a trapping
 potential well outside the horizon that can trap the superradiant modes.
 If there exists a trapping potential well, superradiant amplification
 of the bound state in the potential well will trigger the instability
 of black hole. Otherwise, the black hole is stable although the
 superradiance is present.

 As a matter of convenience, we want to introduce a new radial function as
$\psi=\Delta^{1/2}R$.
 The radial wave equation (5) can be rewritten as
 \begin{eqnarray}
 \frac{d^2\psi}{dr^2}+(\omega^2-V)\psi=0\;,
 \end{eqnarray}
 with the effective potential
 \begin{eqnarray}
 V=\omega^2-\frac{1}{\Delta^2}\left[U
 +\frac{1}{4}\left(\frac{Q^2}{M}-2M\right)^2\right]\;.
 \end{eqnarray}

 Now, we analyze the behavior of the effective potential outside of the
 horizon. From Eq.(20), after some algebra, we can get the derivative of the effective potential as
 \begin{eqnarray}
 V'&=&\frac{1}{\Delta^3}\left(r-\frac{Q^2}{M}\right)
 \left[2\left(r-\frac{Q^2}{M}\right)^2(\omega r-qQ)(2M\omega-qQ)\right.\nonumber\\
 && -2M\mu^2 \left(r-\frac{Q^2}{M}\right)^2(r-2M)
 -l(l+1)\left(r-\frac{Q^2}{M}\right)(r-2M)\nonumber\\
 &&\left. -l(l+1)(r-2M)^2
 \right]\;.
 \end{eqnarray}
 It is convenient to define a new variable
 $z=r-\frac{Q^2}{M}$,
 in terms of which the derivative of the effective potential can be rewritten as
 \begin{eqnarray}
 V'=\frac{z}{\Delta^3}(az^3+bz^2+cz+d)\;,
 \end{eqnarray}
 with
 \begin{eqnarray}
 &&a=2\omega(2M\omega-qQ)-2M\mu^2\;,\\
 &&b=2(2M\omega-qQ)\left(\frac{\omega Q^2}{M}-qQ\right)
 -2M\mu^2\left(\frac{Q^2}{M}-2M\right)-2l(l+1)\;,\\
 &&c=-3l(l+1)\left(\frac{Q^2}{M}-2M\right)\;,\\
 &&d=-l(l+1)\left(\frac{Q^2}{M}-2M\right)^2\;.
 \end{eqnarray}
 The behavior of the effective
 potential can be roughly described by the properties
 of the roots of $V'=0$.

 We now analysis the case of the nonextremal black hole.
 Obviously, $z=0$ is a root of $V'(z)=0$. The other three roots
 are denoted as $\{z_1,z_2,z_3\}$. It is well-known that the relations between
 the roots and the coefficients of the
 cubic equation $az^3+bz^2+cz+d=0$ are given by
 \begin{eqnarray}
 &&z_1+z_2+z_3=-\frac{b}{a}\;,\\
 &&z_1z_2+z_1z_3+z_2z_3=\frac{c}{a}\;,\\
 &&z_1z_2z_3=-\frac{d}{a}\;.
 \end{eqnarray}

 For the scalar field satisfying the superradiant condition
 and the bound state condition, the frequency $\omega$
 of the mode satisfies
 \begin{eqnarray}
 0\leq\omega<\textrm{min}\{qQ/2M, \mu\}\;.
 \end{eqnarray}
 Then, it is obvious that
 \begin{eqnarray}
 a<0\;.
 \end{eqnarray}
 This implies that
 \begin{eqnarray}
 V'(r\rightarrow\infty)\rightarrow 0^-\;.
 \end{eqnarray}
 Note also that
 \begin{eqnarray}
 V(r\rightarrow 2M)\rightarrow-\infty\;,
 \end{eqnarray}
 and
 \begin{eqnarray}
 V(r\rightarrow Q^2/M)\rightarrow-\infty\;.
 \end{eqnarray}
 This implies that there exists at least one maximum point
 in the region $r>2M$ and at least one maximum point in the region
 $Q^2/M<r<2M$. We denote this two maximum points as $z_1$ and
 $z_2$ respectively. Then we have
 \begin{eqnarray}
 z_1>z_2>0\;.
 \end{eqnarray}

 For the nonextremal black hole, $Q^2<2M^2$.
 Then, we have
 \begin{eqnarray}
 c>0\;,\;\;\;d<0\;.
 \end{eqnarray}
 This implies
 \begin{eqnarray}
 z_3<0\;.
 \end{eqnarray}
 This means that in the superradiant regime the effective potential
 $V(r)$ has only one maximum outside the
 event horizon. This implies that there is no trapping potential well outside
 of the event horizon which is separated from the horizon by a potential barrier.

 \begin{figure}
    \includegraphics[width=0.4\textwidth]{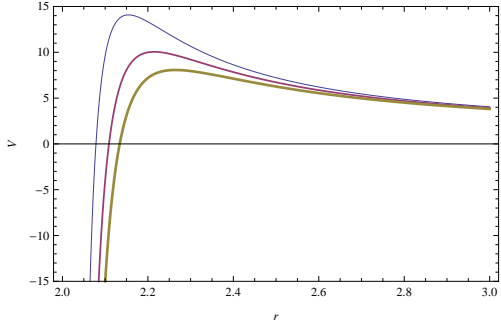}
    \caption{Qualitative shape of the effective potential
    $V$ for different $\omega$. The parameters are given by
    $M=1,Q=1,q=1,\mu=1$ and $l=1$. From top to down, the three curves correspond
    to $\omega=1/3,1/4$ and $1/5$ respectively.}
 \end{figure}

 In FIG. 1, we have plotted the shape of the effective
 potential $V$ given in Eq.(15) for different $\omega$.
 The analytical conclusion for the nonextremal black hole case
 is explicitly shown in this figure.
 The mass of the scalar field
 is never able to generate a potential well outside of the horizon
 to trap the superradiant modes.
 Thus, there are no meta-stable bound states of the
 charged massive scalar field in the superradiant regime.
 In other words, the superradiance in the nonextremal charged stringy
 black hole can not trigger the instability.

 For the case of the extremal black hole, $Q^2=2M^2$.
 The effective potential becomes
 \begin{eqnarray}
 V=\omega^2-\frac{1}{(r-2M)^2}\left[(\omega r-qQ)^2-
 \mu^2r(r-2M)-l(l+1)\right]\;.
 \end{eqnarray}
 The derivative of the effective potential is then given by
 \begin{eqnarray}
 V'=\frac{2}{z^3}(az+b)\;,
 \end{eqnarray}
 with
 \begin{eqnarray}
 &&a=2\omega(2M\omega-qQ)-2M\mu^2\;,\\
 &&b=2(2M\omega-qQ)^2
 -2l(l+1)\;.
 \end{eqnarray}
 The coefficient $a$ is still unchanged, i.e. $a<0$ in the
 superradiant region.
 The root of $V'(z)=0$ is simply given by
 \begin{eqnarray}
 z_0=-\frac{b}{a}\;.
 \end{eqnarray}
 We shall analysis the following two situations in detail.

 Case I: $(2M\omega-qQ)^2<l(l+1)$ or $b<0$

 The effective potential behaves
 \begin{eqnarray}
 V(r\rightarrow 2M)\rightarrow +\infty\;.
 \end{eqnarray}
 But $z_0<0$, i.e. the root of $V'(z)=0$ locates at the
 non-physical regime $r<2M$.
 This implies that there is neither an maximum
 point nor an minimum point outside the horizon.
 In the black-hole exterior, the effective potential will gradually
 bring down to a finite value.

 \begin{figure}
    \includegraphics[width=0.4\textwidth]{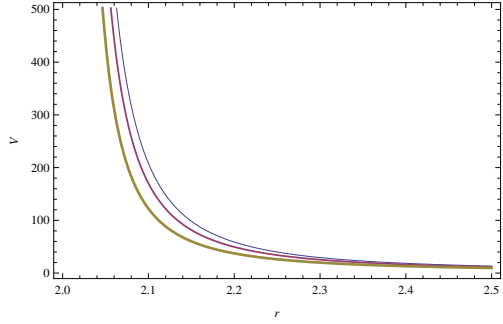}
    \caption{Qualitative shape of the effective potential
    $V$ for different $\omega$. The parameters are given by
    $M=1,Q=\sqrt{2},q=1,\mu=1$ and $l=1$.
    From right to left, the three curves correspond
    to $\omega=1/2,1/3$ and $1/5$ respectively.}
 \end{figure}

 In FIG. 2, we have plotted the shape of the effective potential
 given in Eq.(33) for different $\omega$.
 The parameters are selected to satisfying the
 extremal condition and $b<0$ simultaneously.
 One can see that, outside the horizon,
 there exists neither a potential barrier nor a potential well.
 In this case, the superradiance can not lead to the
 instability.

 Case II: $(2M\omega-qQ)^2>l(l+1)$ or $b>0$

 The effective potential behaves
 \begin{eqnarray}
 V(r\rightarrow 2M)\rightarrow -\infty\;.
 \end{eqnarray}
 But $z_0>0$, i.e. there is a root of $V'(z)=0$ in the region
 of $r>2M$. This implies that the effective potential have only one maximum
 point outside the horizon, i.e. there is only a potential barrier outside
 of the horizon.

 \begin{figure}
    \includegraphics[width=0.4\textwidth]{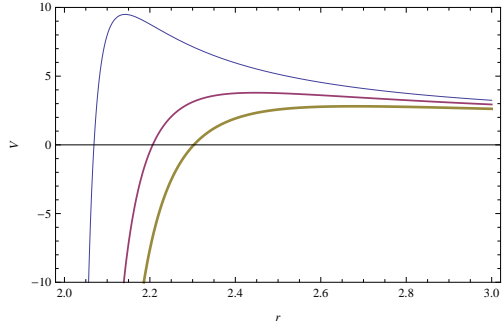}
    \caption{Qualitative shape of the effective potential
    $V$ for different $\omega$. The parameters are given by
    $M=1,Q=\sqrt{2},q=1,\mu=1$ and $l=0$.
    From top to down, the three curves correspond
    to $\omega=1/2,1/3$ and $1/4$ respectively.}
 \end{figure}

 In FIG. 3, we have plotted the effective potential for
 different $\omega$ which satisfy the condition $b>0$.
 The shape of the effective potential in this case is very
 similar to that of the nonextremal black hole. So for the same reason,
 in the present case, the superradiant modes can not be trapped
 as well and the black hole is also stable.

 In summary, we have firstly shown that the classical superradiance
 phenomenon presents in the charged stringy black holes for the charged scalar field\
 perturbations. The superradiant condition is also obtained by analyzing the asymptotic
 solutions near the horizon and at the spatial infinity, which is similar
 to that of RN black hole. Then we investigate the possibility of instability
 triggered by the superradiance. It is shown by analyzing the behavior of the effective
 potential that for both the nonextremal black holes and the extremal black holes
 there is no potential well which is separated from the horizon by a potential barrier.
 Thus, the superradiant modes of charged massive scalar field can not be trapped
 and lead to the instabilities of the black holes.
 This indicates that the extremal and the nonextremal charged black holes in string theory
 are stable against the massive charged scalar field perturbations.

 At last, we should note that although the mass of the scalar field can not
 provide an effective potential well outside the black hole, one can still make the
 black hole unstable by placing a reflecting mirror around the black hole \cite{press,cardoso2004bomb,
 Hod,herdeiro,hod2013prd}. It would be interesting to study the behavior of the scalar field
 in this black hole-mirror system in the future work.

 \section*{ACKNOWLEDGEMENT}
 This work was supported by NSFC, China (Grant No. 11205048).

 \end{document}